\documentstyle[aps,prl,epsf]{revtex}
\def\Tr{{\rm{Tr}}}

\begin{document}
\title{
\centerline{Simulating Quantum Mechanics by Non-Contextual
Hidden Variables} }

\author{Rob Clifton${}^1$}
\address{Departments of Philosophy and History and Philosophy of
Science, \\ 10th floor, Cathedral of
Learning, University of
Pittsburgh, \\ Pittsburgh, PA\ 15260, USA.}
\author{Adrian Kent${}^2$}
\address{
Centre for Quantum Computation,
Clarendon Laboratory, Department of Physics, \\
University of Oxford, Parks Road, Oxford OX1 3PU, U.K. \\
\vskip 5pt
and
\vskip 5pt
Department of Applied Mathematics and
Theoretical Physics, University of Cambridge,\\
Wilberforce Road, Cambridge CB3 0WA, U.K.${}^3$  \\
}

\date{9th August, 1999; revised 10th February, 2000}
\maketitle
\begin{abstract}
No physical measurement can be performed with infinite precision.
This leaves a loophole in the standard no-go
arguments against non-contextual hidden variables.  All such arguments
rely on
choosing special sets of quantum-mechanical observables with measurement
outcomes that cannot
be simulated non-contextually.
As a consequence, these arguments do not exclude
the hypothesis that the class of physical measurements in fact corresponds to a
dense subset of all theoretically possible measurements with
outcomes and quantum probabilities that \emph{can} be recovered from a
non-contextual
hidden variable model.
We show here by explicit construction that there
are indeed such non-contextual hidden variable models, both for
projection valued and positive operator valued measurements.
\vfill
\end{abstract}
\vskip15pt
\noindent
${}^1$ Electronic address: rclifton+@pitt.edu\hfill\break
${}^2$ Electronic address: apak@damtp.cam.ac.uk\hfill\break
${}^3$ Permanent address
\vfill\eject

\section{introduction}\label{introduction}

Bell's theorem (Bell 1964) establishes that
local hidden variable theories are committed
to Bell inequalities violated by quantum correlations.  Since violations of
Bell inequalities can be verified
without requiring that the observables whose correlations figure in
the inequalities be
measured with arbitrarily high precision, Bell's theorem yields a
method of falsifying local hidden variable theories.  Moreover, the
evidence for violations of Bell inequalities very strongly suggests
that such theories have indeed been
falsified (Clauser et al. 1969, Aspect et al. 1981, Tittel et
al. 1998, Weihs et al. 1998). 

We are concerned here, however, not with locality but with
non-contextuality.
Non-contextual hidden variable theories ascribe definite truth values to
projections or, in the case of generalised measurements, any
positive operators, so that the truth values predict the outcome of any
measurement involving the relevant operator and are independent of
the other projections or positive operators involved in
the measurement.  Of course, non-contextual hidden variable
 theories that reproduce the quantum correlations between spatially
 separated systems must, by Bell's theorem, be
non-local.  However, our interest here is in determining whether
\emph{non-relativistic}
 quantum
theory can be simulated classically via non-contextual hidden
variables.  Since non-relativistic classical mechanics does not
presuppose a light cone structure, non-locality is not a meaningful
constraint on hidden variables in this context, and we shall
henceforth ignore
questions of non-locality altogether.

Unlike the arguments against local hidden variables, the known
arguments against non-contextual hidden variables
require observables to be measurable with perfect precision.
These arguments derive from the work of Gleason (1957), 
Bell (1966) and Kochen and Specker (1967). (For more recent
discussions see Redhead 1989, Mermin 1993, Zimba and Penrose 1993, 
Peres 1995, Bub 1997.)

Pitowsky (1983, 1985) argued some time ago that  
these no-go arguments could be evaded by restricting attention
to appropriately chosen subsets of the space of observables.
While ingenious, Pitowsky's models, which are 
constructed via the axiom of choice and the continuum 
hypothesis, have the defect that they rely for their interpretation
on a radically non-standard version of probability theory, 
according to which (for example) the conjunction of two 
probability one events can have probability zero. 

More recently, Meyer (1999) has emphasized that the fact that all
physical measurements are of finite precision leaves a  
loophole in the arguments against non-contextual hidden variables.
One could hypothesize that the class
of possible physical measurements is only a dense subset of the
full set of von Neumann or positive operator valued measurements.
That is, in any given finite precision measurement there is a
fact of the matter, unknown to us, as to which precise measurement
is being carried out, and these realised measurements always belong
to some particular dense subset, which again need not necessarily be
known to us.  Under this hypothesis, the arguments against 
non-contextual hidden variables, which rely
on ascribing definite values to all projections in a real
three-dimensional space (or to certain well chosen finite subsets of
projections), have been shown not to go through (Meyer 1999), nor
can any similar arguments be constructed for either projections or 
positive operator valued measurements in Hilbert
spaces of three or higher dimensions (Kent 1999).
These recent counterexamples rely only on
constructive set theory. 

The aim of the present paper is more ambitious.  We shall show
that all the predictions of
non-relativistic quantum mechanics that are verifiable to within any finite
precision \emph{can} be simulated
classically by non-contextual hidden variable
theories.  That is, there are non-contextual hidden variable
models whose predictions
are practically indistinguishable from those of non-relativistic quantum
mechanics for either projection valued or positive operator
valued measurements.  We give explicit examples, whose construction
requires only constructive set theory and whose interpretation  
needs only elementary (standard) probability theory.  

Before giving details, we should explain why we find the question
interesting.
We have no particular interest in advocating non-contextual hidden variable
theories.  However, we believe that it is important to
distinguish strongly held theoretical beliefs from rigorously
established facts in analysing the ways in which quantum theory is
demonstrably non-classical.
We share, too, with Meyer (1999) another motivation: questions
about the viability of hidden variable
models for a particular quantum process translate into
questions about the classical simulability of some particular aspect
of quantum behaviour, and are interesting independently of the
physical plausibility of the relevant models.
In particular, from the point of
view of quantum computation, the precision attainable in a measurement
is a computational resource.  Specifying infinite precision requires
infinite resources and prevents any useful comparison with discrete
classical computation.  It is interesting to see that,
once the assumption of infinite precision is relaxed, the
outcomes and probabilities of quantum measurements can indeed
be simulated classically and non-contextually.

\section{Outline of Results}\label{outline}

We begin by reviewing the standard theoretical argument against
non-contextual hidden variable models, which is based on infinite
precision von Neumann, i.e. projection valued, measurements
performed on a system
represented by an $n$-dimensional Hilbert space $H_{n}$, where
$n$ is finite.

Determining a unique value for some measured
observable $O$, with spectral projections $\{P_{i}\}$, is equivalent to
distinguishing
exactly one member of the set $\{P_{i}\}$ and assigning it value `1' (to
signify that the corresponding eigenvalue of $O$ is determined
to occur on measurement), while
assigning all the other projections in $\{P_{i}\}$ value `0'.  Let
$\mathcal{P}$ denote a set of
projections, and $\overline{\mathcal{P}}$ denote the set of all
observables whose spectral projections lie in $\mathcal{P}$.  Then, whether
there can be
hidden variables that uniquely determine the measured values of all
the observables in $\overline{\mathcal{P}}$
is equivalent to asking whether there exists a
\emph{truth function} $t:\mathcal{P}\rightarrow \{\mbox{0},\mbox{1}\}$
satisfying:
\begin{equation} \label{eq:eq}
\sum_{i}t(P_{i})=1\ \mbox{whenever}\ \sum_{i}P_{i}=I\ \mbox{and}\
\{P_{i}\}\subseteq \mathcal{P}.
\end{equation}
If $n>2$, and we take
$\overline{\mathcal{P}}$ to be all observables of the
system, it is an
immediate consequence of Gleason's theorem (Gleason 1957) that
$\mathcal{P}$ admits no truth function.
A simpler proof of the relevant part of Gleason's result was given by
Bell (1966), who discussed its implications for
hidden variable models.

Kochen and Specker (1967)
exhibited
a finite set of spin-1 observables
$\overline{\mathcal{P}}$
such that $\mathcal{P}$ admits no truth function, and arguments for
the nonexistence of truth functions on finite sets of projections are often
(summarily) called the
`Kochen-Specker Theorem'.
Simpler examples of finite sets of observables admitting no truth
function have since been given by Peres (1995) and
Zimba and Penrose (1993), among others.

Clearly, then, even before statistical considerations
enter, for a hidden variable theory to remain viable,
its hidden variables cannot determine unique values for all
observables, or even certain finite subsets of observables.
There is a simple intuitive reason why contradictions can arise when
one attempts to construct a truth function on some set of
projections $\mathcal{P}$.
Recall that two projections $P, P'$ are said to be compatible
if $[ P , P' ] = 0$.
Call two resolutions of the identity operator
$\sum_{i}P_{i}=I$ and $\sum_{j}P'_{j}=I$
\emph{compatible} if $[P_{i},P'_{j}]=0$ for all $i,j$.   If the
various
resolutions of the identity generated by $\mathcal{P}$
are all mutually compatible, then closing $\mathcal{P}$ under products of
projections and complements yields a Boolean
algebra under the operations $P\wedge Q=PQ$ and $P^{\perp}=I-P$, and
Boolean algebras always possess truth functions.  
However, in dimension $n>2$, one can
choose $\mathcal{P}$  so that it generates resolutions of the identity
with many projections in common between different \emph{incompatible}
resolutions (in particular, such $\mathcal{P}$'s fail to generate
a Boolean algebra).  This
can drastically reduce the number of `unknowns' relative to equations in
(\ref{eq:eq}) to
the point where they have no solution.

Conversely, one \emph{can}
have a viable hidden variable theory (still assuming infinite
measurement precision) only if one is prepared to assign
a value to a projection that is not simply a function of the hidden
variables, but also of the resolution of the identity that the
projection is considered to be a member of.
Physically, this would mean that the measured value of the
projection would be allowed to depend upon the context in which it is
measured, i.e., on the complete commuting set of
observables that the projection is jointly measured with --- hence the
phrase `contextual hidden variable theory'.  Alternatively, one can
adopt the approach Kochen and Specker themselves advocated, and view
the nonexistence of truth functions as an argument for a
quantum-logical conception of a system's properties.

However, when it comes to considering practical experiments, we are
not actually forced either towards contextual hidden variables or
quantum logic by the Kochen-Specker theorem.
The Pitowsky models (Pitowsky 1983) mentioned above provide 
one possible alternative approach.  Unfortunately,
both the axiom of choice and the continuum hypothesis (or some
weaker axiom in this direction) are needed to define 
Pitowsky's models.  Moreover, the non-standard 
version of probability theory required for their interpretation has
such bizarre properties that we doubt whether these models 
can reasonably be said to constitute a classical explanation or 
simulation of quantum theory.  
At the very least, these features decrease the value of the 
comparison with classical physics and classical computation.

However, the more recent constructive arguments by Meyer (1999) and
Kent (1999) show that one can
always find a subset $\mathcal{P}_{\mathnormal{d}}$ of projections on
$H_{n}$, for any finite $n$, such that $\mathcal{P}_{\mathnormal{d}}$
admits a truth function $t$ and
generates a countable dense set of resolutions of the identity.
By the
latter, we
mean that for any $k$-length resolution $\sum_{i=1}^{k}P_{i}=I$ (where
$k\leq n$) and any $\epsilon>0$, one
can always find
another $k$-length resolution $\sum_{i=1}^{k}P'_{i}=I$ such that
$|P_{i}-P'_{i}|<\epsilon$ for all $i$ and
$\{P'_{i}\}\subseteq\mathcal{P}_{\mathnormal{d}}$.  In particular,
for any self-adjoint operator $O$ and any $\epsilon>0$,
there is an $O'\in\overline{\mathcal{P}}_{\mathnormal{d}}$, with the
same eigenspectrum as $O$, such that the probabilities for
measurement outcomes of $O'$ lie within $\epsilon$ of the
corresponding probabilities for $O$.

Given this, the non-contextual hidden variable theorist is free to adopt
the hypothesis that any \emph{finite} precision
measurement by which we attempt an approximate measurement of an observable
$O$ with spectral decomposition $\sum_{i=1}^{k} a_i P_i$ actually corresponds
to a measurement of some other observable $O'=\sum_{i=1}^{k} a_i
P'_i$ lying in the
precision range and belonging to
$\overline{\mathcal{P}}_{\mathnormal{d}}$.
Moreover, one can allow this observable $O'$ to be specified by the state
and characteristics
of the measuring device alone: $O'$ need not depend
either on the quantum state of the system or on the hidden variables of the
particle being measured.
The hypothesis, in other words, is that when we set up an experiment
to carry out a finite precision measurement there is some fact of the
matter as to which precise observable is actually being measured, and
the measured observable in fact
belongs to $\overline{\mathcal{P}}_{\mathnormal{d}}$.
The precise observable being measured is presumably specified
by hidden variables associated with the measuring apparatus.
It is not known to us in any given experiment; we know only the
precision range.

This, it should be stressed, does {\it not} amount to
reintroducing contextualism into the hidden variable theory.
While the distribution of hidden variables associated with the measured
system needs to be related to the quantum state of that system in
order to recover its quantum statistics, these variables are
supposed to be independent of both the
hidden variables associated with the measuring apparatus and
\emph{its} quantum state; in
particular, they are not correlated with the unknown choice of $O'$.
Note too that the fact that measurements are finite precision is
not sufficient in itself to overthrow the standard analysis.
If we assumed, as above, that some precise $O'$ was always
specified, and we allowed the possibility that $O'$ could be chosen to be any
observable sufficiently close to $O$, the Kochen-Specker theorem would
again threaten.
It is the restriction to
observables in $\overline{\mathcal{P}}_{\mathnormal{d}}$, that has a
set of projections
\emph{admitting} a truth function,
that creates a loophole.

Thus, a finite precision
attempt to measure $O$ can have its outcome specified, in a
noncontextual way, by the value
picked out for $O'$ by the truth function on
$\mathcal{P}_{\mathnormal{d}}$.  Invoking
this toy noncontextual hidden variable model, the practical import
of the theoretical argument against such models
supplied by the Kochen-Specker theorem can be regarded as `nullified' (to
use Meyer's term).

Meyer and Kent leave open the question as to whether
we can actually construct a non-contextual hidden variable model that is
consistent with all the
\emph{statistical} predictions of quantum theory.   Let us call a
collection
of truth functions on a set $\mathcal{P}$ \emph{full} if for any two
distinct $P,P'\in\mathcal{P}$, there exists a truth function $t$ on
$\mathcal{P}$
such that $t(P)\not=t(P')$.  For example, if $\mathcal{P}$ generates
a Boolean algebra, then it will possess a full set of truth
valuations (one for each minimal nonzero projection of $\mathcal{P}$), but
the converse fails.
Unless $\mathcal{P}_{\mathnormal{d}}$
possesses a full set of truth valuations, the previous paragraph's
toy model cannot satisfy the statistical predictions of quantum theory.
For suppose that two distinct $P,P'\in\mathcal{P}_{\mathnormal{d}}$ are
always mapped to
the same truth value under all truth valuations.  Then every set of
hidden variables must dictate the same values for $P$ and $P'$, and
therefore the theory will have to predict the same expectation values
for $P$ and $P'$ in every quantum state of the system.  But this is
absurd: since $P$ and $P'$ are distinct,
there is certainly some quantum state of the system (which we need only
assume is prepared to
within finite precision) in which the expectations of $P$ and $P'$ differ.

There certainly are sets of projections $\mathcal{P}$ that admit truth
valuations, but not a full set (see the `Kochen-Specker diagram' on p. 70 of
Kochen and Specker 1967, involving only 17 projections).  Moreover, it is not
obvious that
any $\mathcal{P}_{\mathnormal{d}}$ with the above properties --- for
example those given by Meyer (1999) and Kent (1999) --- must
necessarily possess a full set of truth valuations.
Hence the unfalsifiability of non-contextual
hidden variables in the face of quantum statistics has yet to be
established.
The goal of the present paper is to establish this unfalsifiability.

In Section \textbf{III}, we shall prove the following result:
\begin{quote}
\textbf{Theorem 1}\ \emph{There exists a set of projections
$\hat{\mathcal{P}}_{\mathnormal{d}}$
on $H_{n}$, closed under products of compatible projections and
complements, that generates a
countable dense set of resolutions of the identity with the property
that no two compatible projections in $\hat{\mathcal{P}}_{\mathnormal{d}}$
are members
of incompatible resolutions.}
\end{quote}
Consider, first, the structure of $\hat{\mathcal{P}}_{\mathnormal{d}}$.
Regard two resolutions of the identity in $\hat{\mathcal{P}}_{\mathnormal{d}}$
as equivalent if they are compatible.  Then since compatible
projections in $\hat{\mathcal{P}}_{\mathnormal{d}}$
can only figure in compatible resolutions, we obtain an equivalence
relation.  To see its transitivity, suppose $\{P_{i}\}$,
$\{P'_{j}\}$ and $\{P'_{j}\}$,
$\{P''_{k}\}$ are compatible pairs of resolutions in
$\hat{\mathcal{P}}_{\mathnormal{d}}$,
fix some arbitrary
indices
$i,j,k$, and consider the
resolution $I=P_{i}P'_{j}+(I-P_{i})P'_{j}+(I-P'_{j})$.  Since
$[I-P'_{j},P''_{k}]=0$, we must have $[P_{i}P'_{j},P''_{k}]=0$, which in
turn entails
\begin{equation}
P''_{k}P_{i}P'_{j}=P_{i}P'_{j}P''_{k}=P_{i}P''_{k}P'_{j}.
\end{equation}
 Summing this equation over $j$ yields $[P''_{k},P_{i}]=0$, and hence
 the resolutions $\{P_{i}\}$,
$\{P''_{k}\}$ must be compatible as well.

It follows that the relation of
compatibility between resolutions of the identity generated by
$\hat{\mathcal{P}}_{\mathnormal{d}}$ partitions that set
into a collection of Boolean algebras that share only the
projections $0$ and $I$ in common.   In particular, the
truth valuations on $\hat{\mathcal{P}}_{\mathnormal{d}}$ are full, since
each of its Boolean subalgebras possesses a full set of truth
valuations, and any collection of assignments of truth values to all
the Boolean
subalgebras of $\hat{\mathcal{P}}_{\mathnormal{d}}$ extends trivially to a
truth valuation on the whole of $\hat{\mathcal{P}}_{\mathnormal{d}}$.

Moreover, it should already be clear that the set of truth valuations on
$\hat{\mathcal{P}}_{\mathnormal{d}}$ will be sufficiently rich to recover
the statistics of any
quantum state by averaging over the values of the hidden variables that
determine
the various truth valuations.  

First, given any state $D$, and nonzero projection
$P\in\hat{\mathcal{P}}_{\mathnormal{d}}$, there will always be many truth
valuations that
map $t(P)=1$, and we may assign the subset of hidden variables for
which $t(P)=1$ measure $\Tr(DP)$.  This prescription also works for
calculating joint probabilities of compatible projections
 $P,P'\in\hat{\mathcal{P}}_{\mathnormal{d}}$, since
 \begin{equation}
 \mbox{Prob}_{D}(P=1,P'=1)=\mbox{Prob}_{D}(PP'=1)=\Tr(DPP'),\
 \mbox{and}\ PP'\in\hat{\mathcal{P}}_{\mathnormal{d}}.
 \end{equation}

So, given any Boolean subalgebra $B$ 
of $\hat{\mathcal{P}}_{\mathnormal{d}}$, we can define a measure
on the hidden variables such that  
 \begin{equation} \label{boolmeas}
 \mbox{Prob}_{D}(P=1)=\Tr(DP),\
 \mbox{for all}\ P\in B. 
 \end{equation}
Since the Boolean subalgebras of $\hat{\mathcal{P}}_{\mathnormal{d}}$ 
are disjoint apart from the elements $0$ and $I$, which have measure
$0$ and $1$ respectively under this definition, the product measure
defined by (\ref{boolmeas}) and 
 \begin{equation}
 \mbox{Prob}_{D}(P=1,P'=1)=\mbox{Prob}_{D}(P=1) \mbox{Prob}_{D}(P'=1),\
 \mbox{for all incompatible}\ P, P'\in\hat{\mathcal{P}}_{\mathnormal{d}}.
 \end{equation}
gives a consistent definition of a measure on the hidden variables
and reproduces all the predictions of quantum mechanics concerning
projection valued measurements. 

To sum up, then,
the existence of the set $\hat{\mathcal{P}}_{\mathnormal{d}}$ defeats
the practical possibility of falsifying
non-contextual hidden variables on either nonstatistical or
statistical grounds.

The same goes for falsifying classical logic.  Following Kochen and
Specker, let
us call a set of
projections
$\mathcal{P}$ a \emph{partial Boolean algebra} if it is
closed under products of compatible projections and complements, and
a truth valuation $t:\mathcal{P}\rightarrow \{\mbox{0},\mbox{1}\}$
will be called a
\emph{two valued
homomorphism} if $t$ also preserves compatible products and complements
(i.e., $t(PP')=t(P)t(P')$ and $t(P^{\perp})=1-t(P)$).  In their
Theorem $0$, Kochen and Specker establish that if
$\mathcal{P}$ possess a full set of two valued
homomorphisms, then $\mathcal{P}$ is imbeddable into a
Boolean algebra.  Since $\hat{\mathcal{P}}_{\mathnormal{d}}$ is closed
under the relevant operations, it is a partial Boolean algebra, and
it is clear that all its truth valuations will in fact be two valued
homomorphisms.  Thus $\hat{\mathcal{P}}_{\mathnormal{d}}$ can always be
imbedded into a Boolean algebra, and this, from a practical point of
view, nullifies any possible argument for a quantum logical
conception of properties.

In Section \textbf{IV},
we rule out falsifications of non-contextual models based on generalized
observables represented by POV measures.  Let $\mathcal{A}$ be a set of
positive
operators on $H_{n}$, and consider all the positive operator (PO)
decompositions of
the identity that $\mathcal{A}$ generates, i.e., decompositions
$\sum_{i}A_{i}=I$ with $\{A_{i}\}\subseteq\mathcal{A}$.  Since
$\sum_{i}A_{i}=I$ does not entail $A_{i}A_{j}=0$ for $i\not=j$, there
can be more members of $\{A_{i}\}$ than the dimension, $n$, of the
space, and the POs in a resolution of the identity need not be
mutually compatible.  Still, we can ask the
analogous question: does there exist a
truth function $t:\mathcal{A}\rightarrow \{\mbox{0},\mbox{1}\}$
satisfying
\begin{equation}
\sum_{i}t(A_{i})=1\ \mbox{whenever}\ \sum_{i}A_{i}=I\ \mbox{and}\
\{A_{i}\}\subseteq \mathcal{A}?
\end{equation}
Moreover: does $\mathcal{A}$ possess enough truth valuations to
recover the statistical predictions, prescribed by any
state $D$, that pertain to the
members of $\mathcal{A}$?  Again, we show the answer is `Yes' for
some sets $\mathcal{A}$ containing countable dense sets of
\emph{finite} PO resolutions of the identity.  By this we mean
(just as in the projective resolution case) that for any
$k$-length PO
decomposition $\sum_{i}^{k}A_{i}=I$ and $\epsilon>0$, there is
another $k$-length PO
decomposition $\sum_{i}^{k}A'_{i}=I$ such that $|A_{i}-A'_{i}|<\epsilon$
for all $i$
and $\{A'_{i}\}\subseteq
\mathcal{A}$.

Note that from the point of view of practically
performable measurements, there is no need to consider infinite PO
resolutions of the identity $\sum_{i}^{\infty}A_{i}=I$ (which, of
course, exist even in finite dimensions).  The reason is
that, for any state $D$ of the system, a POV measurement
amounts to ascertaining the values of the numbers $\Tr(DA_{i})$.
Since these numbers normalize to unity, only a finite subset of
them are measurable to within any finite precision.  Thus any infinite
POV measurement is always practically equivalent to a finite one.
Similarly, any von Neumann measurement of an observable is always
practically equivalent to the measurement of an observable with finite
spectrum.  For this reason, we have not sought to generalize any of
our results to infinite-dimensional Hilbert spaces, though it might
well
be of theoretical interest to do so.

Specifically, in Section \textbf{IV} we
shall establish:
\begin{quote}
\textbf{Theorem 2}\ \emph{There exists a set of POs
$\hat{\mathcal{A}}_{\mathnormal{d}}$
on $H_{n}$ that generates a countable dense set of finite PO
resolutions of the identity  with the property that no two
resolutions share a common
PO.}
\end{quote}
Though this result is weaker than its analogue for projections in
Theorem 1, it still insulates noncontextual models of
POV measurements from falsification.  Because the resolutions
generated by $\hat{\mathcal{A}}_{\mathnormal{d}}$ fail to overlap,
the truth values for POs within a resolution may be set quite
independently of the values assigned to POs in other resolutions.
Thus,
it is clear that there will be sufficiently many such valuations to
recover the statistics of any density operator $D$.  And, as before,
we can suppose that any purported POV measurement, of any length $k$, actually
corresponds to a POV measurement corresponding to a $k$-length PO
resolution (within the precision range of the
measurement) that lies in $\hat{\mathcal{A}}_{\mathnormal{d}}$.

\section{Non-contextual Hidden Variables for PV measures}

Our goal in this section is to
establish Theorem 1.  Let $H_{n}$ be an $n$-dimensional Hilbert space,
and denote
an (ordered) orthonormal basis of $H_{n}$ by
\begin{equation}
\langle e_{i}\rangle = \{ e_{1},e_{2},\ldots,e_{n}\} \, .
\end{equation}
Let $M$ be the metric space whose points are orthonormal
bases,
with the distance between two bases $\langle e_{i}\rangle$ and $\langle
e'_{i}\rangle$
given by $|I-U|$,
where $U$ is the unitary operator mapping $e_{i}\mapsto
e'_{i}$
for all $i$.  Next, consider $U(n)$, the unitary group on $H_{n}$
(endowed
with the operator norm topology), and recall that
$U(n)$
is compact.  Fix a reference point $\langle e_{i}\rangle\in M$, and
consider the
mapping $\varphi:M\mapsto U(n)$
defined by $\varphi(\langle e'_{i}\rangle)=U$, where $U$ is the unitary
operator that maps $e_{i}\mapsto e'_{i}$ for all $i$.
Evidently $\varphi$ is a homeomorphism, thus $M$ is a compact,
complete, separable metric space.

Call two points
$\langle e_{i}\rangle,\langle e'_{i}\rangle\in M$ \emph{totally
incompatible} whenever
\emph{every} projection onto a subspace generated by a nonempty proper
subset of the
vectors $\langle e_{i}\rangle$ is incompatible
with \emph{every} projection onto a subspace generated by a nonempty
proper subset of the vectors
$\langle e'_{i}\rangle$.  We shall need to make use of the following:

\textbf{Lemma 1}: For any finite sequence
\begin{equation}
\langle e_{i}^{(1)}\rangle,\ldots,\langle e_{i}^{(m-1)}\rangle\in M,
\end{equation}
the subset of points, $T^{(m)}$, that are totally incompatible with
all members of the
sequence, is dense in $M$.

\textbf{Proof}:  Let the indices $k$ and $l$ range over the
values of some fixed enumeration of the (proper)
subsets of $\{1,\ldots,n\}$,
let $P_{k}^{(j)}$ denote the projection onto the subspace generated by the
$k$th subset of
$\langle e_{i}^{(j)}\rangle$ (for $j=1,...,m-1$), and for an arbitrary
unlabelled point
$\langle f_{i}\rangle\in M$, let $P_{l}$ be the projection onto the
subspace generated by the $l$th
subset of $\langle f_{i}\rangle$.    Define:
\begin{equation}
I_{kl}^{(j)}\stackrel{\mathrm{def}}{=}\left\{\langle f_{i}\rangle\in M:
[P_{l},P_{k}^{(j)}]\not=0 \right\}.
\end{equation}
Clearly $T^{(m)}$ is just the finite intersection of all the sets of
form $I_{kl}^{(j)}$ over all $j,k,l$.  Now the intersection of any two open
dense sets in $M$ is again an open dense set. 
  So if we can argue that each $I_{kl}^{(j)}$ is both open
and dense, then it will follow that $T^{(m)}$ is dense in $M$.

  So fix $j,k,l$ once and for all.  To see that $I_{kl}^{(j)}$ is open,
  pass to its complement
  $\overline{I}_{kl}^{(j)}$, and consider any Cauchy sequence $\{\langle
  f_{i}^{(p)}\rangle\}_{p=1}^{\infty}\subseteq\overline{I}_{kl}^{(j)}$
  with limit $\langle f_{i}\rangle\in M$.  We must show $\langle
  f_{i}\rangle\in\overline{I}_{kl}^{(j)}$, i.e., that
$[P_{l},P_{k}^{(j)}]=0$.
  By hypothesis, $[P_{l}^{(p)},P_{k}^{(j)}]=0$ for all $p$, and $\langle
  f_{i}^{(p)}\rangle\rightarrow\langle
  f_{i}\rangle$.  Let $U_{p}$ be the unitary operator mapping $f_{i}\mapsto
  f_{i}^{(p)}$ for all $i$.  Then
  $P_{l}^{(p)}=U_{p}P_{l}U_{p}^{-1}$ for all $p$, $U_{p}\rightarrow
  I$ (in operator norm),
  and we have:
   \begin{eqnarray}
  0 & = & \lim_{p\rightarrow\infty}[P_{l}^{(p)},P_{k}^{(j)}]   \\
  & = &
     \lim_{p\rightarrow\infty}\left[U_{p}P_{l}U_{p}^{-1},P_{k}^{(j)}\right]
= [P_{l},P_{k}^{(j)}] \, . \nonumber
 \end{eqnarray}

  To see that $I_{kl}^{(j)}$ is dense, fix an arbitrary point $\langle
  f_{i}\rangle\in M$, and arbitrary $\epsilon>0$.  We must show
  that one can always find a unitary $U$ such that:
  \begin{equation} \label{eq:first}
  |I-U|<\epsilon\ \mbox{and}\ [UP_{l}U^{-1},P_{k}^{(j)}]\not=0;
  \end{equation}
  for then the point $\langle
  Uf_{i}\rangle$ must lie inside $I_{kl}^{(j)}$ and within $\epsilon$
  of $\langle
  f_{i}\rangle$.  Since $j,k,l$ are all fixed, we are free to set $P=P_{l}$ and
  $Q=P_{k}^{(j)}$ for simplicity, bearing in mind that $P,Q\not= 0
  {\rm~or~} I$.
  To establish (\ref{eq:first}), then, all we need to show is that assuming
  \begin{equation} \label{eq:second}
  |I-U|<\epsilon\ \ \Rightarrow\ \  [UPU^{-1},Q]=0
  \end{equation}
  leads to a contradiction.

  First, we dispense with the case $P=Q$.  Consider the
  one parameter group of unitaries $U_{t}\equiv e^{itH}$ where $H$ is
  self-adjoint.  Since $Q\not=0,I$, we may suppose that $[H,Q]\not=0$.  By
  (\ref{eq:second}), $U_{t}QU_{-t}$
  and $Q$ commute for all sufficiently small $t$, in which case we may write
  \begin{equation} \label{eq:forgot!}
  Q =A_{t}+B_{t},\ \  U_{t}QU_{-t}
  =A_{t}+C_{t},
  \end{equation}
  where $A_{t}$, $B_{t}$, and $C_{t}$ are pairwise
  orthogonal projections.  Then,
  \begin{equation} \label{eq:wow}
  0 =\lim _{t\rightarrow 0}
  |U_{t}QU_{-t}-Q| =\lim _{t\rightarrow 0} |B_{t}-C_{t}|.
  \end{equation}
  By (\ref{eq:wow}), we may choose a $\delta>0$ so that
$|B_{t}-C_{t}|<\frac{1}{2}$ for all
  $t<\delta$.  If $B_{t}$ were nonzero for some $t<\delta$, then we
  could choose a
  unit vector $e$ in the range of $B_{t}$, and in that case we would
  have
  $\|(B_{t}-C_{t})e\|=\|e\|=1$.  However, this
  contradicts the fact that $|B_{t}-C_{t}|<\frac{1}{2}$.  Thus,
  in fact $B_{t}=0$ for all $t<\delta$, and by symmetry $C_{t}=0$ for all
  $t<\delta$.
  Hence, for all $t<\delta$, $U_{t}QU_{-t}=A_{t}=Q$, i.e., $[U_{t},Q]=0$.
  However, since $\lim_{t\rightarrow 0}t^{-1}(U_{t}-I)= iH$, any operator
  that commutes with all $U_{t}$ in a neighborhood of the identity
  must commute with $H$.  Thus $[H,Q]=0$, contrary to hypothesis.

  Next, consider the case of general $P$ and $Q$ ($\not=0,I$).  Since
$[P,Q]=0$,
  we may write $Q =A+B$, $P=A+C$,
where $A,B$ and $C$ are pairwise
  orthogonal projections.  Without loss of generality, we may assume
  $A\not=0$ (i.e., $PQ\not=0$); for if not, then we may replace $P$ by
  $I-P$ (in order to guarantee $PQ=A\not=0$), and under
  that replacement (\ref{eq:second}) continues to hold.  Similarly,
  we may assume that $A+B+C\not=I$; for if not, we could replace $Q$
  by $I-Q$.  Since neither $A+B+C$ nor $A$ equals $0$ or $I$, there is a
  self-adjoint $H'$ such that $[H',B]=[H',C]=0$ but $[H',A]\not=0$.
  Defining $U_{t}\equiv e^{itH'}$, we have that
  $[U_{t}PU_{-t},Q]=[U_{t}AU_{-t},A]$  for all $t$.  Thus
  (\ref{eq:second}) implies:
  \begin{equation} \label{eq:third}
  |I-U_{t}|<\epsilon\ \ \Rightarrow\ \  [U_{t}AU_{-t},A]=0,
  \end{equation}
which, in turn, entails the contradiction $[H',A]=0$ by the argument of the
previous paragraph (with $H'$ in place of $H$, and $A$ in place of $Q$).
\emph{QED}.

 \textbf{Proposition 1}: There is a countable dense subset of
$M$
whose members are pairwise totally incompatible.

\textbf{Proof}:
Since $M$ is
  separable, it possesses a countable dense set
  $T=\{\langle e_{i}^{(1)}\rangle,\ldots,\langle
e_{i}^{(m-1)}\rangle,\ldots\}$.
  If, in moving along
  this sequence, some point
  $\langle e_{i}^{(m)}\rangle$ were found to be not totally incompatible
  with all
previous members of the sequence, then, by Lemma 1, we could always discard
$\langle e_{i}^{(m)}\rangle$  and replace it with
a new
point $\langle \hat{e}_{i}^{(m)}\rangle\in T^{(m)}$.   The replacement point
$\langle \hat{e}_{i}^{(m)}\rangle$ \emph{will}
be totally incompatible with all previous members, and (since $T^{(m)}$ is
dense)
it can always be chosen to lie within a distance $2^{-m}$ of $\langle
e_{i}^{(m)}\rangle$.
Moving down the sequence $T$, and replacing points in this way as many times
as needed, we obtain a new
countably infinite sequence $\hat{T}$ that is pairwise totally incompatible.
And
$\hat{T}$
is itself dense.  For in every ball $B(p, \epsilon) =\{ q : d (p, q) <
\epsilon\} $ around any point $p\in M$ there must
be infinitely many members of the dense set $T$.  If one of those
members did not need replacing, then clearly $B(p, \epsilon )$ contains an
element of $\hat{T}$.  But this must also be true if all of them needed
replacing, because the replacement points
$\langle\hat{e}_{i}^{m}\rangle$
lie closer and closer to $\langle e_{i}^{m}\rangle$ as
$m\rightarrow\infty$.  \emph{QED}.

Now we can complete the proof of Theorem 1.
Let $\{\langle \hat{e}_{i}^{(1)}\rangle,\langle
\hat{e}_{i}^{(2)}\rangle,\ldots,\}$ be the dense subset of Proposition 1.
Let the $k$ in $\hat{P}_{k}^{(m)}$ (the projection onto the span of the
$k$th subset of $\langle \hat{e}_{i}^{(m)}\rangle$) now range over an
enumeration
of \emph{all} the subsets of
$\{1,\ldots,n\}$ (including the empty set, corresponding to $k=0$ and
$\hat{P}_{0}^{(m)}=0$, and the entire set, corresponding to $k=2^{n}$ and
$\hat{P}_{2^{n}}^{(m)}=I$).
Define:
\begin{equation}
B_{m}\stackrel{\mathrm{def}}{=}
\left\{\hat{P}^{(m)}_{k}:k=1,\ldots,2^{n}\right\},\ \
\hat{\mathcal{P}}_{\mathnormal{d}}\stackrel{\mathrm{def}}{=}
\bigcup_{m=1}^{\infty}B_{m}.
\end{equation}
Clearly each $B_{m}$ is a maximal Boolean algebra.
Moreover, Proposition 1 assures us that any
two (nontrivial) compatible projections
$P,P'\in\hat{\mathcal{P}}_{\mathnormal{d}}$ must lie in the same $B_{m}$,
as well as all the resolutions of the identity in which $P,P'$
figure.  Thus compatible projections in
$\hat{\mathcal{P}}_{\mathnormal{d}}$ only appear in compatible
resolutions.  Trivially, $\hat{\mathcal{P}}_{\mathnormal{d}}$ is
closed under compatible products and complements, since each $B_{m}$
is so closed, and the projections contained in different $B_{m}$'s
(excepting $0$
and $I$) are all incompatible.  Finally, since $\{\langle
\hat{e}_{i}^{(1)}\rangle,\langle
\hat{e}_{i}^{(2)}\rangle,\ldots,\}$ is dense in $M$, it is immediate that
$\hat{\mathcal{P}}_{\mathnormal{d}}$ generates a dense set of
resolutions of the identity, and Theorem 1 is proved.
Note, finally, that by specifying from the outset a particular countable
dense subset
of $U(n)$, all our arguments in the proof can be made constructively.
(In particular, the argument for (9) could have been given directly,
rather than via  a reductio ad absurdum from (10).)

\section{Non-Contextual Hidden Variables for POV measures}

We turn, next, to establish Theorem 2.
Let $\mathcal{O}$ be the set of all operators on $H_{n}$, endowed with
the operator norm topology, and let $\mathcal{O}^{+}$
denote the PO's on $H_{n}$, a closed subset of $\mathcal{O}$.    Fix, once and
for all, some
basis $B\subseteq H_{n}$, and consider the matrix representations of
all operators in $\mathcal{O}$ relative to $B$.
Define $\mathcal{Q}_{\mathnormal{C}}$ to
be the set of all operators with complex rational matrix entries
(relative to $B$), and $\mathcal{Q}_{\mathnormal{C}}^{+}$ to
 be the subset of positive operators
therein.  Clearly $\mathcal{Q}_{\mathnormal{C}}$ is dense in $\mathcal{O}$.
It follows that
$\mathcal{Q}_{\mathnormal{C}}^{+}$  is dense in $\mathcal{O}^{+}$.
For consider any $A\geq 0$. Then there is an
$X\in \mathcal{O}$ such that $A=X^{\ast}X$.  Since there is a sequence
$\{X_{m}\}_{m=1}^{\infty}\subseteq
\mathcal{Q}_{\mathnormal{C}}$ converging to $X$,
$\{X_{m}^{\ast}X_{m}\}_{m=1}^{\infty}$
is a sequence in
$\mathcal{Q}_{\mathnormal{C}}^{+}$ converging to $A$.
Next, define $\mathcal{A}$
to be the subset of positive operators in
$\mathcal{Q}_{\mathnormal{C}}^{+}$ whose (complex rational) matrix entries
are all \emph{nonzero}.  To see that $\mathcal{A}$
is also dense in $\mathcal{O}^{+}$, it suffices to observe that it is dense in
$\mathcal{Q}_{\mathnormal{C}}^{+}$.  So consider any $A\in
\mathcal{Q}_{\mathnormal{C}}^{+}$, choose any
$A'\in\mathcal{A}$, and let $\{t_{m}\}_{m=1}^{\infty}$ be a sequence of
positive
rationals tending to $0$.  Then clearly $A+t_{m}A'\rightarrow A$, each
$A+t_{m}A'\in \mathcal{Q}_{\mathnormal{C}}^{+}$ (since the positive
operators form a convex
cone, and the rationals a field), and at most $n^{2}$ of the operators
$\{A+t_{m}A'\}_{m=1}^{\infty}$ can have a zero matrix entry (since the
operators $A$ and $A'$ are fixed).

We now need to establish:

\textbf{Lemma 2}: For any $k$, $\mathcal{A}$ generates a dense set of
$k$-length PO resolutions
of the identity.

\textbf{Proof}:
Let $\sum_{i=1}^{k}A_{i}=I$ be any $k$-length PO
resolution of the identity, and fix $\epsilon>0$.  Choose a rational
$r\in(0,\epsilon)$ and set $\delta=r/(5+k)>0$.  Then, since
$\mathcal{A}$ is dense in $\mathcal{O}^{+}$, we may choose $k$ POs
$\{A'_{i}\}_{i=1}^{k}\subseteq\mathcal{A}$ such that
$|A_{i}-A'_{i}|<\delta$ for all $i$.  With $H=\sum_{i=1}^{k}A'_{i}$,
observe that
\begin{equation}
-\delta I<-|A'_{i}-A_{i}|I\leq A'_{i}-A_{i}\leq |A'_{i}-A_{i}|I<\delta I,
\end{equation}
 which, summed over
$i=1,\ldots,k$, yields
\begin{equation} \label{eq:help}
 -\delta kI<H-I<\delta kI.
 \end{equation}

 Next, introduce the new
positive operators:
\begin{equation} \label{eq:reveals}
A''_{i}=
(1+k\delta)^{-1}\left[A'_{i}+ t_i \left((1+k\delta)I-H\right)\right],
\end{equation}
for all $i$.
Here the $t_i$ are positive rationals obeying $\sum_{i=1}^{k}t_i = 1$ and
$  t_i < 2/k $, and are
chosen so that all the matrix entries of all the $A''_i$ are
nonzero.  Such a choice can always be made, since for each fixed $i$ there are
only finitely many choices one can make for $t_i$ (in fact, at most
$n^{2}$) such that a matrix entry of $A''_i$ can vanish.
Each $A''_{i}$ is positive, since $A'_{i}\geq 0$,
$(1+k\delta)I-H>0$ (by (\ref{eq:help})), and the positive operators form a
convex cone.  Each $A''_{i}$ is also complex rational (hence lies in
$\mathcal{Q}_{\mathnormal{C}}^{+}$), since
each $A'_{i}$ and, therefore, $H$ is,
as is $I$ (obviously), and all of $k,\delta, t_{i}$ are rational.
Thus,
$\{A''_{i}\}_{i=1}^{k}\subseteq\mathcal{A}$.
Summing
(\ref{eq:reveals}) over $i=1,\ldots,k$ reveals that
$\sum_{i=1}^{k}A''_{i}=I$.  Finally, note that since
\begin{equation}
|H-I|=\inf\{\lambda>0:-\lambda I\leq H-I\leq\lambda I\},
\end{equation}
(\ref{eq:help}) entails $|H-I|\leq\delta k$.  And, since
$\sum_{i=1}^{k}A_{i}=I$,
$|A_{i}|\leq 1$ for all $i$.  With latter inequalities,
the triangle inequality, $  t_i < 2/k $, and the inequalities of the previous
paragraph, we obtain:
\begin{eqnarray}
   |A_{i}-A''_{i}|  \leq  & \left|(1+k\delta)^{-1}\left[A'_{i}+t_i
\left((1+k\delta)I-H\right)\right]-
A_{i}\right|  \\
   \leq &
    (1+k\delta)^{-1}|A'_{i}-(1+k\delta)A_{i}|
+t_i (1+k\delta)^{-1}|(1+k\delta)I-H| \nonumber \\
    \leq &
    (1+k\delta)^{-1}|A'_{i}-A_{i}- k\delta
    A_{i}| +t_i (1+k\delta)^{-1}|I-H+k\delta I| \nonumber \\
\leq &
    (1+k\delta)^{-1}(\delta +k\delta)+t_i (1+k\delta)^{-1}(\delta
    k+k\delta) \nonumber \\
    \leq & \delta(5+k)/(1+k\delta)< \delta(5+k) = r<\epsilon \nonumber
    \, ,
\end{eqnarray}
  for all $i$.  \emph{QED}.

Now since $\mathcal{A}$ consists only of complex rational POs, it is
countable.  The set of all finite subsets of a countable set is itself
countable.  Thus $\mathcal{A}$ can include at most countably many finite PO
decompositions of the identity. Let the variable $m$
range over an enumeration of these resolutions, denoting the $m$th
resolution, which will have some length $k_{m}$,
by
$\{A_{i}^{(m)}\}_{i=1}^{k_{m}}$.
For each $m$, define the unitary operator $U_{m}$ to be given by the diagonal
matrix ${\rm diag} ( e^{i\theta_{m}} , 1 , \ldots , 1 )$ relative to the
basis $B$, where
\begin{equation} \label{eq:expressions}
 \sin \theta_{m} = ( \pi /4 )^m,\ \
   \cos \theta _{m} = (1 - ( \pi /4 )^{2m} )^{1/2} \, ,
   \end{equation}
and for definiteness we take the positive square root.
For each $m$, define a new $k_{m}$-length PO resolution
$\{\hat{A}_{i}^{(m)}\}_{i=1}^{k_{m}}$ by
\begin{equation} \label{eq:wow2}
 \hat{A}_{i}^{(m)}=
U_{m}A_{i}^{(m)} U_{m}^{-1},\ \mbox{for all}\ i=1,\ldots,k_{m}.
\end{equation}
Then we have:

\textbf{Proposition 2}: No two of the PO resolutions
$\{\hat{A}_{i}^{(m)}\}_{i=1}^{k_{m}}$ (for different $m$) share a common PO.

\textbf{Proof}:
Assuming that for some $i,j$ and $m,p$ we have
$\hat{A}_{i}^{(m)}=\hat{A}_{j}^{(p)}$, we must show that
$m=p$.  By (\ref{eq:wow2}),
\begin{equation} \label{eq:matrix}
U_{m}A_{i}^{(m)} U_{m}^{-1}=U_{p}A_{j}^{(p)} U_{p}^{-1}.
\end{equation}
Let $(x_{ab})$ and $(y_{ab})$ be the nonzero complex rational matrix
coefficients (relative to $B$) of $A_{i}^{(m)}$ and $A_{j}^{(p)}$,
respectively.  Using the definition of $U_{m}$, (\ref{eq:matrix})
entails, in particular, that
$x_{12}e^{i\theta_{m}}=y_{12}e^{i\theta_{p}}$.  Since $x_{12}\not=0$,
we may write:
\begin{equation}\label{eq:yy}
e^{i(\theta_{m}-\theta_{p})}=c,
\end{equation}
where $c$ is a complex
rational. Equating the real parts of (\ref{eq:yy}), one obtains
\begin{equation}  \label{eq:square}
\cos\theta_{m}\cos\theta_{p}=r-\sin\theta_{m}\sin\theta_{p},
\end{equation}
where $r$ ($=\Re(c)$) is rational.  Squaring both sides of
(\ref{eq:square}), inserting the expressions (\ref{eq:expressions}),
and rearranging, yields:
\begin{equation}
2r\left(\frac{\pi}{4}\right)^{m+p}-\left(\frac{\pi}{4}\right)^{2p}-
\left(\frac{\pi}{4}\right)^{2m}=r^{2}-1.
\end{equation}
The transcendentality of $\pi$ requires that the coefficient of any
given power of $\pi$ in this equation must be zero.  This cannot
happen unless $m=p$, since if $m\not =p$, then $\pi^{m+p}$, $\pi^{2p}$,
and $\pi^{2m}$ are three distinct powers of $\pi$, and the latter two powers
occur with nonzero coefficient.\ \emph{QED}.

We finish with the argument for Theorem 2.  All that remains to show
is that the
countable collection of finite PO
resolutions contained in
\begin{equation}
\hat{\mathcal{A}}_{\mathnormal{d}}\stackrel{\mathrm{def}}{=}
\bigcup_{m=1}^{\infty}\{\hat{A}_{i}^{(m)}\}_{i=1}^{k_{m}}
\end{equation}
 is dense. Fix $k$ and let
$\{A_{i}\}_{i=1}^{k}$ be an arbitrary $k$-length
resolution.  By Lemma 2, we can find $k$-length resolutions in
$\mathcal{A}$ arbitrarily close to $\{A_{i}\}_{i=1}^{k}$.  Since the
unitary
operators
defined by (\ref{eq:expressions})
satisfy $U_{m}\rightarrow I$, by Proposition 2 we can find $k$-length
resolutions in
$\hat{\mathcal{A}}_{\mathnormal{d}}$ arbitrarily close to any $k$-length
resolution in  $\mathcal{A}$.
This completes the proof.

\section{Discussion}

To many physicists, the fact that quantum theory can most elegantly
be expressed in a radically non-classical language makes
explicit demonstrations of its non-classicality
essentially redundant.  This paper is addressed
not to them, but to those who, like us, are interested
in what we can establish for certain on the question.
As we have already stressed, we hold no particular brief for
non-contextual hidden variable theories.  However, our conclusion, in the
light of the results of Meyer (1999) and Kent (1999) and
the constructions above, is that there is no truly compelling
argument establishing that non-relativistic quantum mechanics
describes classically inexplicable physics.
Only when quantum theory and relativity are combined can
really compelling arguments be mounted against the possibility
of classical simulation, and then --- so far as is presently
known --- only against the particular class of simulations
defined by local hidden variable theories.

One feature of our toy models is that the 
non-contextual hidden variables at any given time define
outcomes only for one measurement, not 
for a sequence of measurements at separate times. 
This is adequate for our purposes, since we are interested only in 
establishing the point that, despite the Kochen-Specker theorem,
non-contextual hidden variables can
simulate the measurement process in non-relativistic 
quantum mechanics.  

As our model has no dynamics, 
it cannot supply a proper account of processes extended in time.  
However, our discussion could be extended to a treatment
of sequential projective or positive operator valued measurements 
assuming there is no intervening evolution, simply by assuming 
that the hidden variables, like the state vector,
undergo a discontinuous change after a measurement,
so that the probability distribution of the post-measurement
hidden variables corresponds to that defined by the new state vector. 
If such postulates were adopted, they would, in the spirit of the
hiddern variables program, need to be seen as approximations to be 
justified by a more fundamental theory with complete dynamics.
A complete theory would also, of course, need to describe 
successive measurements in which the intervening evolution of
the quantum state is non-trivial. 

Constructing a dynamical non-contextual hidden variable theory with
these properties goes beyond our ambitions here, and it should be stressed that
we have not shown that our toy models, or any non-contextual hidden
variable theories, can in fact be extended to a viable dynamical
non-contextual hidden variable theory.  
It is sufficient for our
argument here to note that Kochen-Specker constraints {\it per se} give
no reason to think that such a theory cannot be constructed.
In fact, though, we see no argument against 
the possibility of such a theory, given that 
it is possible to give a precise collapse dynamics for quantum
states (Ghirardi et al. 1986).

From another viewpoint, it might be argued that our proofs and constructions
establish more than is strictly necessary.
While we have defined a hidden variable simulation
of quantum theory by constructing a dense subset of the set of projective
decompositions with the property that no two compatible projections
belong to incompatible resolutions, this last property is not
necessary.  To give a trivial example, one could also define a
simulation on the basis of a dense set of projections exactly one of
which belongs to two (or more)
incompatible resolutions.  Similar comments apply to the
analysis for positive operator decompositions.

Our response would simply be that the constructions we give do
the job.   They are certainly not unique, and there may well
be other constructions which differ in interesting ways, or
even perhaps define more natural non-contextual
hidden variable models.

A related argument that might also be made is that the
constructions of Meyer (1999) and Kent (1999) already imply the
simulability of quantum theory by non-contextual hidden variables,
since they each describe a truth valuation $t$ on a dense set in which both
truth values occur densely (i.e., for any projection or positive
operator mapped to 1 by $t$, there is another one arbitrarily close to
it mapped to 0).
One could imagine a model in which the particle, confronted
by a measuring apparatus set with a particular precision range,
first calculates the approximate probability
that it should produce a `1' by evaluating the expectation in its
quantum state of some projection randomly chosen from within the
precision range,
and then goes through some deterministic algorithm to {\it choose}
a (generally different) projection from the range for which it
will reveal its pre-determined measurement value.  In this picture,
it is left to the deterministic algorithm to simulate quantum
statistics by choosing projections of value `0' or `1' in the
right proportions.  

Such models cannot be logically excluded, but
they seem to us overly baroque: hidden variable theories in which the effects
of the system and apparatus hidden variables can be separated
seem to us cleaner and simpler than theories which rely on some
conspiratorial interaction between those variables at the point
of measurement.

\vskip10pt \centerline{\bf Acknowledgements} 
\vskip5pt 
We thank J. Finkelstein, N.D. Mermin and an anonymous referee for helpful
comments.  A.K. was supported by a Royal Society University Research
Fellowship and thanks the Oxford Centre for Quantum Computation for
much appreciated hospitality.  
\vskip5pt

\vskip.3in
\leftline{\bf References}\par
\vskip.1in

Aspect, A. et al. 1981
Experimental tests of realistic local theories via Bell's theorem.
{\it Phys. Rev. Lett.}  {\bf 47}, 460--463.
\vskip.1in
 Bell, J.S. 1964  On the Einstein-Podolsky-Rosen Paradox.
{\it Physics}  {\bf 1}, 195--200.
\vskip.1in
 Bell, J. S. 1966
On the problem of hidden variables in quantum mechanics.
{\it Rev. Mod. Phys.}  {\bf 38}, 447--452.
\vskip.1in
 Bub, J. 1997 
{\it Interpreting the Quantum World}, Ch. 3.
Cambridge: Cambridge Univ. Press.  
\vskip.1in
 Clauser, J. F. et al. 1969
Proposed experiment to test local hidden-variable theories.
{\it Phys. Rev. Lett.}  {\bf 23}, 880--884.
\vskip.1in
Ghirardi, G. et al. 1986
Unified Dynamics for Microscopic and Macroscopic Systems 
{\it Phys. Rev. D}, {\bf 34}, 470-491. 
\vskip.1in
 Gleason, A. M. 1957 
Measures on the closed subspaces of a Hilbert space.
{\it J. Math. Mech.} {\bf 6}, 885--893.
\vskip.1in
 Kent, A. 1999 
Non-contextual hidden variables and physical measurements.
{\it Phys. Rev. Lett.} {\bf 83} 3755-3757. 
\vskip.1in
 Kochen, S. and Specker, E. P. 1967
The problem of hidden variables in quantum mechanics.
{\it J. Math. Mech.} {\bf 17}, 59--87.
\vskip.1in
 Mermin, N. D. 1993
Hidden variables and the two theorems of John Bell.
{\it Rev. Mod. Phys.} {\bf 65}, 803--815.
\vskip.1in
 Meyer, D. 1999
Finite precision measurement nullifies the Kochen-Specker theorem.
{\it Phys. Rev. Lett.} {\bf 83} 3751-3754
\vskip.1in
 Peres, A. 1995
{\it Quantum Theory:  Concepts and Methods}, Ch. 7, pp. 187-211. 
Boston: Kluwer.
\vskip.1in
 Pitowsky, I. 1983
Deterministic model of spin and statistics.
{\it Phys. Rev. D} {\bf 27}, 2316--2326.
\vskip.1in
 Pitowsky, I. 1985
Quantum mechanics and value definiteness.
{\it Phil. Sci.} {\bf 52} 154--156.
\vskip.1in
 Redhead, M. 1989 
{\it Incompleteness, Nonlocality, and Realism},
Ch. 5.  Oxford: Clarendon Press.  
\vfill\eject
 Tittel, W. et al. 1998
Violation of Bell inequalities by photons more than 10 km apart.
{\it Phys. Rev. Lett.} {\bf 81}, 3563-3566.
\vskip.1in
 Weihs, G. et al. 1998
Violation of Bell's inequality under strict Einstein locality conditions.
{\it Phys. Rev. Lett.} {\bf 81}, 5039-5043.
\vskip.1in
 Zimba, J. and Penrose, R. 1993
On Bell non-locality without probabilities: more curious geometry.
{Stud. Hist. Philos. Sci.} {\bf} 24, 697-720.

\end{document}